\documentclass{article}
\usepackage{graphicx}
\usepackage{amsmath}
\usepackage{amssymb}

\usepackage{amsmath}

\newcommand{\bc}{\begin{center}}
\newcommand{\ec}{\end{center}}
\newcommand{\beq}[0]{\begin{equation}}
\newcommand{\eeq}[0]{\end{equation}}
\newcommand{\bea}{\begin{eqnarray}}
\newcommand{\eea}{\end{eqnarray}}

\newcommand{\stheta}{\sin^22\theta_{13}}
\newcommand{\nova}{NO$\nu$A}
\newcommand{\dm}[1]{\Delta m^2_{#1}}
\newcommand{\PCPC}{
\setlength{\unitlength}{1pt}
\begin{picture}(20,7)
\put(0,0){$P_{CP}$}
\setlength{\unitlength}{1cm}
\end{picture}}
\newcommand{\PCPV}{
\setlength{\unitlength}{1pt}
\begin{picture}(20,7)
\put(5,-2){\line(2,1){14}}
\put(0,0){$P_{CP}$}
\setlength{\unitlength}{1cm}
\end{picture}}
\newcommand{\reu}{{\nu_e\rightarrow\nu_\mu}}
\newcommand{\reub}{{\bar{\nu}_e\rightarrow\bar{\nu}_\mu}}
\newcommand{\ruu}{{\nu_\mu\rightarrow\nu_\mu}}
\newcommand{\ruub}{{\bar{\nu}_\mu\rightarrow\bar{\nu}_\mu}}

\newcommand{\ie}{{\it i.e.}}

\begin{document}
 
\baselineskip 18pt
\begin{flushright}
TUM-HEP-580/05
\end{flushright}

\vspace*{5mm}

\begin{center}
{\Large {\bf Future Precision Neutrino Oscillation Experiments \\
             and Theoretical Implications\footnote{Talk presented at the Nobel Symposium 129: 
             Neutrino Physics, Enk\"oping, Sweden, 2004}
}}
 \vskip5mm M. Lindner\footnote{Email address: lindner@ph.tum.de},

Physik Department, Technische Universit\"at M\"unchen,
D-85748 Garching bei M\"unchen, Germany

\bigskip

\begin{abstract}
\noindent
Future neutrino oscillation experiments will lead to precision 
measurements of neutrino mass splittings and mixings. The flavour 
structure of the lepton sector will therefore at some point become 
better known than that of the quark sector. This article discusses 
the potential of future oscillation experiments on the 
basis of detailed simulations with an emphasis on experiments 
which can be done in about ten years. In addition, some theoretical
implications for neutrino mass models will be briefly discussed.

\end{abstract}
\end{center}

\vspace*{6mm}
\section{Introduction}

The observation of atmospheric neutrino oscillations by the 
SuperKamiokande experiment~\cite{Toshito:2001dk} triggered a 
remarkable discovery phase. The initial evidence turned into a 
solid proof of neutrino flavour conversions as well as of the 
L/E dependence as required by oscillations. The solar neutrino 
problem has also been resolved in the last years. The Gallex 
experiment~\cite{Hampel:1998xg} detected initially a rate effect 
which implied flavour conversion on the basis of solar models.
The SNO experiment proved then model independent 
neutrino flavour transitions~\cite{AMcDonald,Ahmad:2002jz}. 
The initially allowed parameter islands were cleaned up by the 
KamLAND experiment, which demonstrated finally with reactor 
anti-neutrinos~\cite{ASuzuki,Eguchi:2002dm} that the so-called 
LMA-solution is correct. Altogether the existing experimental 
results fit now very nicely into a picture with three massive 
neutrinos, which corresponds to the simplest scenario for three 
generations. The only exemption is the disputed LSND 
result~\cite{Church:2002tc}, which would have far reaching 
consequences if it were confirmed, but this possibility will be 
ignored here. The oscillations of three neutrino generations 
involve then two mass-squared 
differences $\dm{12}\simeq\dm{sol.}$ and $\dm{23}\simeq\dm{atm.}$, 
three mixing angles, $\theta_{12}$, $\theta_{23}$, and $\theta_{13}$, 
and a CP-violating phase $\delta$. Atmospheric neutrino data~\cite{YSuzuki} 
and the first results from the K2K long-baseline accelerator 
experiment~\cite{YSuzuki} determine  
$\Delta m_{23}^2 = (2.2^{+0.6}_{-0.4})\times10^{-3}$~eV$^{2}$ 
and $\theta_{23}\approx45^{\circ}$~\cite{YSuzuki,MCGG}, 
whereas solar neutrino data~\cite{solar,SNOsalt}, combined with 
the results from the KamLAND reactor experiment~\cite{ASuzuki} 
lead to $\Delta m_{12}^{2} = (8.2^{+0.3}_{-0.3}) \times 10^{-5}$~eV$^2$ 
and $\tan^2\theta_{12} = 0.39^{+0.05}_{-0.04}$~\cite{MCGG}.
The results can now approximatively be summarized by two independent 
two flavour oscillations where the solar and atmospheric $\dm{}$ 
values are roughly now known.

The key parameter for genuine three flavour effects is the 
mixing angle $\theta_{13}$ which is so far only known to be small 
from the CHOOZ~\cite{Apollonio:1998xe,Apollonio:2002gd} and 
Palo Verde~\cite{Boehm:1999gk} experiments. The current bound for 
$\theta_{13}$ depends on the value of the atmospheric mass 
squared difference and it gets rather weak for 
$\dm{31} \lesssim 2\times 10^{-3}$~eV$^2$. However, in that region 
an additional constraint on $\theta_{13}$ from global solar neutrino 
data becomes important~\cite{fit}. At the current best 
fit value of $\dm{31} = 2.2\times 10^{-3}$~eV$^2$ we have at 
$3\sigma$ the bound $\sin^2\theta_{13} \le 0.041$~\cite{MCGG}.
There is no reason why $\theta_{13}$ should vanish and one should 
expect therefore $\theta_{13}$ to be finite. 

One might think that neutrino oscillations are in future less 
interesting, since it will lead only to parameter improvements
of the leading $2\times 2$ oscillations and maybe a finite parameter 
value of $\theta_{13}$. However, such a view misses completely 
the fact that the neutrino sector is, unlike the quark sector, 
not obscured by hadronic uncertainties. The precision to which 
the underlying flavour information is determined will therefore 
only be limited by the ultimate experimental precision. If
high precision measurements are possible, then they will be very 
sensitive tests of flavour models and related topics, like the
unitarity of three flavours. Genuine three flavour oscillation 
effects occur only for a finite value of $\theta_{13}$ and establishing 
a finite value of $\theta_{13}$ is therefore one of the next milestones 
in neutrino physics. Leptonic CP violation is another three flavour 
effect which can only be tested if $\theta_{13}$ is finite. 
The usual see-saw scenario includes besides $\delta$ 
in addition two further Majorana CP phases in the light neutrino sector, 
as well as other CP phases in the heavy Majorana sector, which are 
involved in leptogenesis. In general the heavy and light CP phases 
are not connected, but most flavour models create relations
between these two sectors, relating thus low energy leptonic CP violation
to leptogenesis and mass models. Precision measurements of neutrino
oscillations test therefore very interesting questions of particle 
physics which are connected to the origin of flavour and to 
phenomenological consequences of flavour.
There is thus a very strong motivation to establish first in 
the next generation of experiments a finite value of $\theta_{13}$ 
in order to aim in the long run at a measurement of leptonic CP 
violation~\cite{Freund:2000ti,Freund:2001ui,Huber:2002rs,Lindner:2002vt,Huber:2002mx}.

\section{Three neutrino oscillation in matter}
\label{sec:osc}

An effective two flavour treatment is insufficient for future oscillation 
experiments and matter effects must be included in addition. The
generalization of the oscillation 
formulae in vacuum to $N$ neutrinos leads to the probabilities for 
flavour transitions $\nu_{f_l} \rightarrow \nu_{f_m}$ given by
\beq
P( \nu_{f_l} \rightarrow \nu_{f_m} ) 
= \underbrace{\delta_{lm} - 4\sum_{i>j} \mathrm{Re} J_{ij}^{f_l
f_m}\sin^2\Delta_{ij}}_{{\mathbf\PCPC}}~ \underbrace{- 2\sum_{i>j} 
\mathrm{Im} J_{ij}^{f_l f_m}\sin 2\Delta_{ij}}_{{\mathbf\PCPV}}
\label{eq:Nosc}
\eeq
where the shorthands $J_{ij}^{f_l f_m} := U_{li}U_{lj}^*U^*_{mi}U_{mj}$
and $\Delta_{ij} := \frac{\Delta m^2_{ij} L}{4E}$ have been used. These
generalized vacuum transition probabilities depend on all combinations 
of quadratic mass differences $\Delta m^2_{ij}=m_i^2-m_j^2$ as well as 
on different products of elements of the leptonic mixing matrix $U$.
We will assume a three neutrino framework, i.e. $1\leq i,j \leq 3$ 
and $U$ is a $3\times 3$ mixing matrix parameterized in the standard way 
\beq
U  =
\left(\begin{array}{ccc} 
c_{12}c_{13} & s_{12}c_{13} & s_{13}e^{-i\delta}\\
- s_{12}c_{23} - c_{12}s_{23}s_{13}e^{i\delta} & 
 c_{12}c_{23} - s_{12}s_{23}s_{13}e^{i\delta} & s_{23}c_{13}\\
s_{12}s_{23} - c_{12}c_{23}s_{13}e^{i\delta} 
& -c_{12}s_{23} - s_{12}c_{23}s_{13}e^{i\delta}
& c_{23}c_{13}\\ 
\end{array} \right)~,
\label{eq:MNS}
\eeq
where $c_{ij}=\cos(\theta_{ij})$ and $s_{ij}=\sin(\theta_{ij})$.
$U$ contains three leptonic mixing angles and one Dirac-like 
leptonic CP phase $\delta$. Note that the most general mixing matrix 
for three Majorana neutrinos contains two further Majorana-like 
CP phases, but it can easily be seen that these extra phases do 
not enter in the above oscillation formulae.
Disappearance probabilities, \ie\ the transitions 
$\nu_{f_l} \rightarrow \nu_{f_l}$, do not even depend on $\delta$,
since $J_{ij}^{f_l f_l}$ is only a function of the modulus of 
elements of $U$. Appearance probabilities, like 
$\nu_{e} \rightarrow \nu_{\mu}$ are therefore the place where
leptonic CP violation can be studied. Eq.~(\ref{eq:Nosc}) contains 
a CP conserving part $\PCPC$ and a CP violating part $\PCPV$, and both 
terms depend on the CP phase $\delta$. An obvious extraction strategy 
for CP-violation would thus be to look at CP asymmetries~\cite{Dick:1999ed}.
Note, however, that the beams of a long baseline experiment traverse 
the Earth and the presence of matter violates CP by itself. This
implies modifications of eq.~(\ref{eq:Nosc}) and it makes a measurement 
of leptonic CP violation more involved. 

The general oscillation formulae in vacuum, eq.~(\ref{eq:Nosc}), lead 
to well known, but rather lengthy trigonometric expressions for the 
oscillation probabilities in vacuum. These expressions become even longer 
and do not exist in closed form when arbitrary matter corrections are 
taken into account. For effectively constant matter densities, which is 
often a good assumption, the problem simplifies somewhat, but the general 
oscillation probabilities are still very lengthy. The Hamiltonian describing 
three neutrino oscillation in matter can then be written in flavour basis as
\beq
H =
\frac{1}{2E_\nu} U
\left(\begin{array}{ccc} 
m_1^2 & 0 & 0 \\ 
0 & m_2^2 & 0 \\ 
0 & 0 & m_3^2  
\end{array} 
\right) U^T
+
\frac{1}{2E_\nu} 
\left( \begin{array}{ccc} 
A+A' & 0 & 0\\ 
~0~ & ~A'~ & ~0~\\ 
~0~ & ~0~ & ~A'~ 
\end{array} 
\right) ~.
\label{eq:Hccnc}
\eeq
The first term describes oscillations in vacuum in flavour basis.
The quantities $A$ and $A'$ in the second term are given by the 
charged current and neutral current contributions to coherent 
forward scattering in matter. The charged current contribution 
is given by
\beq
A = \pm~ \frac{2\sqrt{2}G_F Y \rho E_\nu}{m_n} = 2VE_\nu~,
\eeq
where $G_F$ is Fermi's constant, $Y$ is the number of electrons per 
nucleon, $m_n$ is the nucleon mass and $\rho$ is the matter density. 
A is positive for neutrinos in matter and anti-neutrinos in 
anti-matter, while it is negative for anti-neutrinos in matter 
and neutrinos in anti-matter. The flavour universal neutral current 
contributions $A'$ lead to an overall phase which does not enter the 
transition probabilities. The over-all neutrino mass scale $m_1^2$ can 
be written as a term proportional to the unit matrix and can similarly 
be removed, such that only $\dm{21}$ and $\dm{31}$ remain in the
first term of eq.~(\ref{eq:Hccnc}). After re-diagonalization of the 
Hamiltonian in constant matter density one finds that matter effects 
lead in a very good approximation to an $A$-dependent parameter mapping 
in the 1-3 subspace which can be written as~\cite{Freund:2001pn}
\bea
\sin^2 2\theta_{13,m} &=& \frac{\sin^2 2\theta_{13}}{C_{\pm}^2} ~, 
\label{eq:map1}\\
\dm{31,m} &=& \dm{31}C_{\pm} ~,\\
\dm{32,m} &=& \frac{\dm{31}~(C_{\pm}+1)+ A}{2}~, \\
\dm{21,m} &=& \frac{\dm{31}~(C_{\pm}-1)- A}{2}~.
\label{eq:map4}
\eea
The index $m$ denotes effective quantities in matter, where 
\beq
C^2_{\pm} = \left(\frac{A}{\dm{31}} 
-\cos 2\theta \right)^2 +\sin^2 2\theta~.
\label{eq:Cpm}
\eeq
Note that $A$ in $C_{\pm}$ can change its sign and the mappings for 
neutrinos and anti-neutrinos are therefore different, resulting
in different effective mixings and masses. This is an important 
effect, which will allow detailed tests of coherent forward 
scattering of neutrinos in matter. Note that oscillations in matter 
depend unlike vacuum oscillations on the sign of $\dm{31}$. This 
allows to extract the $sign(\dm{31})$ via matter effects.

Inserting the parameter mappings eqs.~(\ref{eq:map1})-(\ref{eq:map4})
into the full oscillation formulae leads still to quite lengthy expressions 
for the oscillation probabilities in matter, where it is not easy to 
oversee all effects. It is therefore instructive to simplify the problem 
further such that a qualitative analytic understanding of all 
effects becomes possible, while quantitative statements should be evaluated 
numerically with the full expressions. The key for further simplification 
is to expand the oscillation probabilities in small quantities. These 
expansion parameters are $\alpha=\dm{21}/\dm{31} =${\cal O}($10^{-2}$) and 
$\sin^2 2\theta_{13} \leq 0.16$. The matter effects can be parameterized 
by the dimensionless quantity $\hat A=A/\dm{31}=2VE/\dm{31}$, where 
$V=\sqrt{2}G_F n_e$. The oscillation probabilities for all channels 
can in this way be significantly simplified~\cite{Akhmedov:2004ny}.
Using $\Delta \equiv \Delta_{31}$, the leading terms 
for $P(\nu_\mu \rightarrow \nu_\mu)$ and $P(\nu_e \rightarrow \nu_\mu)$ can, 
for example, be written as~\cite{Freund:2001ui,Cervera:2000kp,Freund:2001pn}
\begin{eqnarray}
& & \hspace*{-8mm}P(\nu_\mu \rightarrow \nu_\mu)  \approx \nonumber \\
& & 
1 - \cos^2 \theta_{13} \sin^2 2\theta_{23} \sin^2 {\Delta}
+ 2~{\alpha}~  \cos^2 \theta_{13} \cos^2 \theta_{12} \sin^2
2\theta_{23} {\Delta} \cos{\Delta},
\label{eq:Pdis}\\
\nonumber \\
& & \hspace*{-8mm}P(\nu_e \rightarrow \nu_\mu) \approx 
{\sin^2 2\theta_{13}}\sin^2 \theta_{23}~
 ~\frac{\sin^2({(1\!\!-\!\hat{A})}{\Delta})}{{(1\!\!-\!\hat{A})^2}} \nonumber\\
& & \pm~
{\sin\delta}\cdot {\sin 2\theta_{13}}~{\alpha}~ \sin 2\theta_{12} \cos\theta_{13}
   \sin 2\theta_{23}\sin({\Delta})\frac{
\sin({\hat{A}}{\Delta})\sin({(1\!\!-\!\hat{A})}{\Delta})}{{\hat{A}(1\!\!-\!\hat{A})}}
\nonumber \\
& &+ \cos\delta\cdot {\sin 2\theta_{13}}~ {\alpha}~  \sin 2\theta_{12} \cos\theta_{13}  
\sin 2\theta_{23}\cos({\Delta})\frac{\sin({
\hat{A}}{\Delta})\sin({(1\!\!-\!\hat{A})}{\Delta})}{{\hat{A}(1\!\!-\!\hat{A})}}
\nonumber \\
& &+~  {\alpha^2}~ \sin^2 2\theta_{12} \cos^2 \theta_{23} 
\frac{\sin^2({\hat{A}}{\Delta})}{{\hat{A}^2}}~,
\label{eq:Pap}
 \end{eqnarray}
where ``$+$'' in eq.~(\ref{eq:Pap}) stands for neutrinos and ``$-$'' for 
anti-neutrinos. The most important feature of eq.~(\ref{eq:Pap}) is that all 
interesting effects in the $\reu$ transition depend crucially on
$\theta_{13}$. The size of $\sin^2 2\theta_{13}$ determines thus if the 
total transition rate, matter effects, effects due to the sign of $\dm{31}$ 
and CP violating effects are measurable. This is the reason why the size of
$\theta_{13}$ is one of the most important questions for future oscillation 
experiments.

Before we discuss some features of eqs.~(\ref{eq:Pdis}) and (\ref{eq:Pap})
in more detail, we would like to comment on the underlying assumptions 
and the reliability of these equations. 
First eqs.~(\ref{eq:Pdis}) and (\ref{eq:Pap}) are an expansion in 
terms of the small quantities $\alpha$ and $\sin 2\theta_{13}$.
Higher order terms are suppressed at least by another power of 
one of these small parameters and these corrections are thus typically
at the percent level. Note that the expansion in $\alpha$ is actually
an expansion in the solar and not the atmospheric frequency. The expansion 
does therefore not break down at the first atmospheric oscillation 
maximum, i.e. at $\Delta\simeq 1$, but at much larger baselines before 
the first (sub-dominant) solar oscillation maximum, i.e. at 
$\alpha\Delta\simeq 1$. The latter condition gives an upper bound for 
the baseline where eqs.~(\ref{eq:Pdis}) and (\ref{eq:Pap}) are good
approximations
\beq
L \lesssim 8000\,\mathrm{km} \left(\frac{E_\nu}{\mathrm GeV}\right)
\left(\frac{10^{-4}{\mathrm eV}^2}{\dm{21}}\right)  ~,
\label{eq:range}
\eeq
while the first oscillation maximum sits at $\alpha\cdot L\simeq L/30$.
Eqs.~(\ref{eq:Pdis}) and (\ref{eq:Pap}) are therefore excellent 
approximations at and well beyond the first oscillation maximum of 
long baseline experiments. The matter corrections in eqs.~(\ref{eq:Pdis}) 
and (\ref{eq:Pap}) are derived for constant average matter density
which is a good approximation.

Note that all quantitative results which will be presented are based on 
numerical simulations in matter. The results are therefore not affected
by any approximation. Eqs.~(\ref{eq:Pdis}) and (\ref{eq:Pap}) will only 
be used to understand the problem analytically, which is extremely 
helpful in order to oversee the multi-dimensional parameter space. 

In addition to long baseline experiments, reactor experiments with 
identical near and far detectors have an excellent potential for 
precise measurements. The near detector is used to eliminate many 
common systematical errors and the far detector is located typically 
at a baseline of a few kilometer. For these short baselines matter 
effects can be ignored and one finds to second order in the small 
quantities $\sin 2\theta_{13}$ and $\alpha$ for the oscillation probability 
\beq
1 - P_{\bar{e} \bar{e}} = 
\sin^2 2 \theta_{13} \, \sin^2 \Delta_{31} +  \alpha^2 \, \Delta_{31}^2 \, 
\cos^4 \theta_{13} \, \sin^2 2 \theta_{12} \,.
\label{eq:reactor}
\eeq
At the first atmospheric oscillation maximum, $\Delta_{31}$ is approximately 
$\pi/2$ and $\sin^2 \Delta_{31}$ is close to one, which means that the 
second term on the right-hand side of this equation can be neglected for
$\sin^2 2 \theta_{13} \gtrsim 10^{-3}$. The reactor measurement is dominated 
in this case at short baselines by the product of $\sin^2 2\theta_{13}$ 
and $\sin^2 \Delta_{31}$, which must be measured as deviation from one. 
Eq.~(\ref{eq:reactor}) implies that correlations and degeneracies play 
essentially no role in reactor experiments. The behavior in the 
$\sin^2 2 \theta_{13}$-$\Delta m_{21}^2$-plane will also be different 
since eq.~(\ref{eq:reactor}) is essentially independent of $\Delta m_{21}^2$. 
A reactor experiment will improve the global parameter determination in two ways. 
First, a direct, essentially uncorrelated and clean measurement for 
$\theta_{13}$~\cite{Minakata:2002jv} can be obtained which can be 
used to disentangle the long baseline results. Secondly, the reactor 
measurement can replace the cross-section suppressed anti-neutrino running 
of the accelerator experiments, leading to statistical improvements in the 
neutrino measurements~\cite{Huber:2003pm}. 


\section{Correlations and degeneracies}
\label{sec:CD}

Eqs.~(\ref{eq:Pdis}) and (\ref{eq:Pap}) exhibit certain parameter 
correlations and degeneracies, which play an important role in the 
analysis of long baseline experiments, and which would be hard to understand
in a purely numerical analysis of the high dimensional parameter 
space. The most important properties are:
\begin{itemize}
\item
Eqs.~(\ref{eq:Pdis}) and (\ref{eq:Pap}) depend only on the product 
$\alpha\cdot \sin 2\theta_{12}$ or equivalently 
$\dm{21}\cdot \sin 2\theta_{12}$. This are the 
parameters related to solar oscillations which will be taken
as external input. Note that the product is better determined than 
the product of the individual measurements of $\dm{21}$ and 
$\sin 2\theta_{12}$. 
\item
Next we observe in eq.~(\ref{eq:Pap}) that the second and third 
term contain both a factor $\sin(\hat A\Delta)$, while the last 
term contains a factor $\sin^2(\hat A\Delta)$. Since 
$\hat A\Delta= VL/2$, we find that these factors depend only on 
$L$, resulting in a ``magic baseline'' \cite{Huber:2003ak}
when $VL_{magic}=2\pi$, 
where $\sin(\hat A\Delta)$ vanishes. At this magic baseline only 
the first term in eq.~(\ref{eq:Pap}) survives and 
$P(\nu_e \rightarrow \nu_\mu)$ does no longer depend on $\delta$, 
$\alpha$ and $\sin 2\theta_{12}$. This is in principle very 
important, since it implies that $\sin^2 2\theta_{13}$ can be 
determined at the magic baseline from the first term of 
eq.~(\ref{eq:Pap}) whatever the values and errors of $\delta$, 
$\alpha$ and $\sin 2\theta_{12}$ are. For the given matter density 
of the Earth we find $L_{magic}= 2\pi/V \simeq 8100~\mathrm {\rm km}$
which fits nicely into the Earth. This value is quite amazing, since 
$V$ is given in terms of completely unrelated constants of nature 
like $G_F$. 
\item
Next we observe that only the second and third term of 
eq.~(\ref{eq:Pap}) depend on the CP phase $\delta$, and 
both terms contain a factor $\sin 2\theta_{13}\cdot\alpha$, 
while the first and fourth term of eq.~(\ref{eq:Pap})
do not depend on the CP phase $\delta$ and contain factors of  
$\sin^2 2\theta_{13}$ and $\alpha^2$, respectively.
The extraction of CP violation is thus always suppressed by 
the product $\sin 2\theta_{13}\cdot\alpha$ and the CP violating 
terms are obscured by large CP independent terms if 
either $\sin^2 2\theta_{13} \ll \alpha^2$ or
$\sin^2 2\theta_{13} \gg \alpha^2$. The relative contribution of the
CP phase $\delta$ to the probability is thus largest for 
$\sin^2 2\theta_{13} \simeq 4\theta^2_{13} \simeq \alpha^2$.
\item
Another observation is that the last term in eq.~(\ref{eq:Pap}),
which is proportional to $\alpha^2=(\dm{21})^2/(\dm{31})^2$,
dominates in the limit of tiny $\sin^2 2\theta_{13}$. The error 
of $\dm{21}$ limits therefore for small $\sin^2 2\theta_{13}$ the 
parameter extraction. This last term implies a finite transition 
probability even for $\theta_{13}=0$. Observing 
${\nu_e\rightarrow\nu_\mu}$ or ${\nu_\mu\rightarrow\nu_e}$
appearance transitions does therefore not necessarily establish 
a finite value of $\theta_{13}=0$ in a three flavour framework.
\item
Eqs.~(\ref{eq:Pdis}) and (\ref{eq:Pap}) suggest that transformations 
exist which leave these equations invariant. Therefore degeneracies, 
\ie\ parameter sets having identical oscillation probabilities for a 
fixed $L/E$ are expected. 
An example of such an invariance is given by a simultaneous replacement
of neutrinos by anti-neutrinos and $\dm{31}\rightarrow -\dm{31}$. 
This is equivalent to changing the sign of the second term of 
eq.~(\ref{eq:Pap}) and replacing $\alpha \rightarrow -\alpha$ and
$\Delta \rightarrow -\Delta$, while $\hat A \rightarrow \hat A$.
It is easy to see that eqs.~(\ref{eq:Pdis}) and (\ref{eq:Pap}) are
unchanged, but this is not a degeneracy, since neutrinos and 
anti-neutrinos can be distinguished experimentally. 
\item
The first real degeneracy~\cite{Fogli:1996pv,Barger:2001yr} can be seen 
in the disappearance probability eq.~(\ref{eq:Pdis}), which is invariant 
under the replacement $\theta_{23}\rightarrow \pi/2 - \theta_{23}$.
Note that the second and third term in eq.~(\ref{eq:Pap}) are not really
invariant under this transformation, but this change in the sub-leading 
appearance probability can approximately be compensated by tiny
parameter shifts. This implies that the degeneracy can in principle 
be lifted with high precision measurements in the disappearance channels.
\item
The second degeneracy can be found in the appearance probability 
eq.~(\ref{eq:Pap}) in the 
($\delta-\theta_{13}$)-plane~\cite{Koike:2000jf, Burguet-Castell:2001ez}. 
In terms of $\theta_{13}$ (which is small) and $\delta$ the four 
terms of eq.~(\ref{eq:Pap}) have the structure  
\beq
P(\nu_e \rightarrow \nu_\mu) \approx 
\theta^2_{13}\cdot F_1 +
\theta_{13}\cdot (\pm~\sin\delta F_2 + \cos\delta F_3) +
F_4~,
\label{eq:degth13}
\eeq
where the quantities $F_i$, $i=1,..,4$ contain all the other parameters. 
The requirement $P(\nu_e \rightarrow \nu_\mu)=const.$ leads for both 
neutrinos and anti-neutrinos to parameter manifolds of degenerate
or correlated solutions. Having both neutrino and anti-neutrino beams, 
the two channels can be used independently, which is equivalent to 
considering simultaneously eq.~(\ref{eq:degth13}) for $F_2\equiv 0$ 
and $F_3\equiv 0$. The requirement that these probabilities are now 
independently constant, \ie\ $P(\nu_e \rightarrow \nu_\mu)=const.$ 
for $F_2\equiv 0$ and $F_3\equiv 0$, leads to more constraint manifolds 
in the ($\delta-\theta_{13}$)-plane, but some degeneracies still survive.
\item
The third degeneracy~\cite{Minakata:2001qm} is given by the fact 
that a change in sign of $\dm{31}$ can essentially be compensated 
by an offset in $\delta$. Therefore we note again that the transformation 
$\dm{31} \rightarrow -\dm{31}$ leads to $\alpha \rightarrow -\alpha$,
$\Delta \rightarrow -\Delta$ and $\hat A \rightarrow -\hat A$. 
All terms of the disappearance probability, eq.~(\ref{eq:Pdis}), are 
invariant under this transformation. The first and fourth term in 
the appearance probability eq.~(\ref{eq:Pdis}), which do not depend
on the CP phase $\delta$, are also invariant. The second and third 
term of eq.~(\ref{eq:Pdis}) depend on the CP phase and change by the 
transformation $\dm{31} \rightarrow -\dm{31}$. The fact that these 
changes can be compensated by an offset in the CP phase $\delta$ 
is the third degeneracy.
\item
Altogether there exists an eight-fold degeneracy 
~\cite{Barger:2001yr}, as long as only the $\ruu$, $\ruub$,
$\reu$ and $\reub$ channels and one fixed $L/E$ are considered. 
However, eqs.~(\ref{eq:Pdis}) and (\ref{eq:Pap}) also imply 
that the degeneracies can be broken by using in a suitable way 
information from different $L/E$ values. This can be achieved 
in total event rates by changing or combining different
$L$ or $E$~\cite{Burguet-Castell:2002qx,Barger:2002rr,Huber:2005ep}, 
but it can in principle 
also be done by using information in the event rate spectrum of a
single baseline $L$, which requires detectors with very good energy 
resolution~\cite{Freund:2001ui}. Another strategy to break the 
degeneracies is to include further oscillation channels in the 
analysis (``silver channels'')~\cite{Burguet-Castell:2002qx,Donini:2002rm}.
\end{itemize}
The discussion of this section shows the strength of the analytic 
approximations, which allow to understand the complicated parameter 
interdependence. It also helps to optimally plan experimental 
setups and to find strategies to resolve the degeneracies.

\section{The potential of future neutrino oscillation experiments}

Triggered by the spectacular results in neutrino physics during 
the last ten years, several new experimental projects are under 
way in this field. It is therefore interesting to investigate 
where we should stand in the determination of neutrino oscillation 
parameters in ten years from now. It is also interesting to look
further and to estimate the ultimate precision which could be 
obtained. 

\begin{figure}[!ht]
\begin{center}
\includegraphics[width=10.2cm]{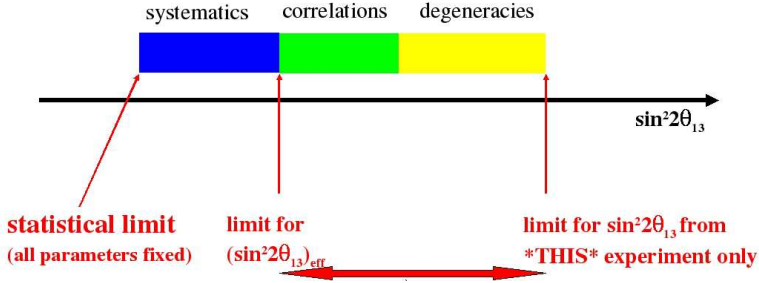}%
\end{center}
\caption{The precision for $\sin^22\theta_{13}$ is shown in 
colored bands, where the left edge of the blue/dark 
grey band shows the initial value which is obtained if 
only statistics is considered. The right edge of the blue/dark 
grey band is the result after the systematic errors are included.
This is the principal precision of the experiment. However, the 
sensitivity for $\sin^22\theta_{13}$ deteriorates further due
to parameter correlations and parameter degeneracies. The final 
value is the right edge of the yellow/light grey band. The range
covered by the green and yellow bands can lead to remarkable 
synergies when this experiment is combined with another experiment
with similar precision, but different parameter dependence.}
\label{fig:ferrors}
\end{figure}

The precision of quantities like $\sin^22\theta_{13}$ which is found
form the simulation of experiments will be presented in a way 
shown in fig.~\ref{fig:ferrors}. The bands show how the 
initial value, which is given by statistics alone (left edge of blue/dark 
grey band) deteriorates by systematic errors, by parameter correlations
(e.g. with the unknown or partly known CP phase) and parameter 
degeneracies (due to trigonometric ambiguities). It is important 
to note that a given experiment (or combination of experiments) 
typically measures some parameter combination with a precision which 
is considerably better than the final limit. This precision of the
experiment is shown in fig.~\ref{fig:ferrors} as the right edge 
of the blue/dark grey band. This precision might be called 
$\left(\sin^22\theta_{13}\right)_{eff}$, since it expresses the 
precision if all other unknown parameters are fixed and no errors 
are included. However, if one properly extracts a 
limit of $\sin^22\theta_{13}$ with all unknowns properly taken into 
account, then one ends up at the right edge of the yellow/light grey 
band. Distinguishing in this way between the precision and the 
sensitivity is quite useful, since it also shows the room for 
improvement by combinations with other similarly precise
experiments with other parameter dependence.

\subsection{Next generation experiments}

Future oscillation experiments can be 
grouped according to their time scale of operation. The K2K
experiment is already running and it tests the leading atmospheric
oscillation already now. Next come the MINOS and CNGS projects 
which are already operating or under construction, respectively. 
Therefore we include in our study the conventional beam experiments
MINOS~\cite{Ables:1995wq}, and the CNGS experiments 
ICARUS~\cite{Aprili:2002wx} and OPERA~\cite{Duchesneau:2002yq}. 
We include also the subsequent superbeam experiments J-PARC to 
SuperKamiokande (T2K)~\cite{Itow:2001ee} and NuMI off-axis 
(\nova)~\cite{Ayres:2002nm}, as well as new reactor neutrino 
experiments~\cite{reactors} with a near and far detector. 
The main characteristics of these experiments are summarized in 
tab.~\ref{tab:experiments}.
For the reactor experiments we use the Double-CHOOZ proposal
(D-CHOOZ)~\cite{doublechooz} as initial stage setup with roughly 
$6\times 10^4$ events, and an optimized setup called Reactor-II, 
with a slightly longer baseline and $6\times 10^5$ events. Such a 
configuration could be realized at several other sites under 
discussion~\cite{reactors}. 
The results presented in the following are based on Ref.~\cite{Huber:2004ug},
where more details on the analysis can be found. The simulations of 
the experiments as well as the statistical analysis is performed with 
the GLoBES software package~\cite{globes}.

\begin{center}
\begin{table}[htb]
\begin{center}
\begin{tabular}{l@{\hspace{-0.8ex}}rrrl}
\hline
 Label &  $L\, [\mathrm{km}]$ &  $\langle E_\nu \rangle$ 
& $t_{\mathrm{run}}$ & 
channel \\
\hline
\multicolumn{5}{l}{ \bf{Conventional beam experiments:}}\\
 MINOS\ & $735$ & 
 $3 \,\mathrm{GeV}$ & $5 \, \mathrm{yr}$ & 
 $\nu_\mu \!\to\! \nu_{\mu,e}$\\
 ICARUS &  $732$ &  
 $17\,\mathrm{GeV}$  & 
 $5 \, \mathrm{yr}$ & 
 $\nu_\mu \!\to\! \nu_{e,\mu,\tau}$\\
 OPERA &   $732$ &  
 $17\,\mathrm{GeV}$ & 
 $5 \, \mathrm{yr}$ & 
 $\nu_\mu \!\to\! \nu_{e,\mu,\tau}$\\
\multicolumn{5}{l}{\bf{Off-axis superbeams:}} \\
 T2K &  $295$ & 
 $0.76 \, \mathrm{GeV}$ & 
 $5 \, \mathrm{yr}$ & 
 $\nu_\mu \!\to\! \nu_{e,\mu}$\\
 \nova\ & 
 $812$ & 
 $2.22 \, \mathrm{GeV}$ & 
 $5 \, \mathrm{yr}$ & 
 $\nu_\mu \!\to\! \nu_{e,\mu}$\\
\multicolumn{5}{l}{
\bf{Reactor experiments:}} \\
 D-CHOOZ & 
 $1.05$ & 
 $\sim 4 \, \mathrm{MeV}$ & 
 $3 \, \mathrm{yr}$ & 
 $\nu_e \!\to\! \nu_e$\\
 Reactor-II & 
 $1.70$ & 
 $\sim 4 \, \mathrm{MeV}$ & 
 $5\,\mathrm{yr}$ & 
 $\nu_e \!\to\! \nu_e$\\
\hline
\end{tabular}
\end{center}
\caption{Characteristics of the considered experiments.}
\label{tab:experiments}
\end{table}
\end{center}


A first interesting question concerns improvements of 
$\Delta m^2_{31}$ and $\sin^2\theta_{23}$. In tab.~\ref{tab:atm} we
show the precision which can be obtained in the future in comparison 
to the current precision, as obtained from a global fit to 
SuperKamiokande (SK) atmospheric and K2K long-baseline 
data~\cite{fit}. The last row is the precision which
can be obtained by combining all experiments. We observe from these numbers,
that the accuracy on $\Delta m^2_{31}$ can be improved by one order of
magnitude, whereas the accuracy on $\sin^2\theta_{23}$ will be improved only
by a factor two.

\begin{table}[htb]
\centering
\begin{tabular}{lrr}
\hline
 & $|\Delta m^2_{13}|$  & $\sin^2\theta_{23}$ \\
\hline
current     & 88\% & 79\% \\
\hline
MINOS+CNGS  & 26\% & 78\% \\
T2K         & 12\% & 46\% \\
\nova\        & 25\% & 86\% \\
\hline
Combination &  9\% & 42\% \\
\hline
\end{tabular}
\caption{Precision for $|\Delta m^2_{31}|$ and $\sin^2\theta_{23}$ at
  $3\sigma$ for the values $\Delta m^2_{31} = 2\times 10^{-3}$~eV$^2$,
  $\sin^2\theta_{23} = 0.5$.}
\label{tab:atm}
\end{table}

\begin{figure}[htb]
\centering
\includegraphics[width=0.75\textwidth]{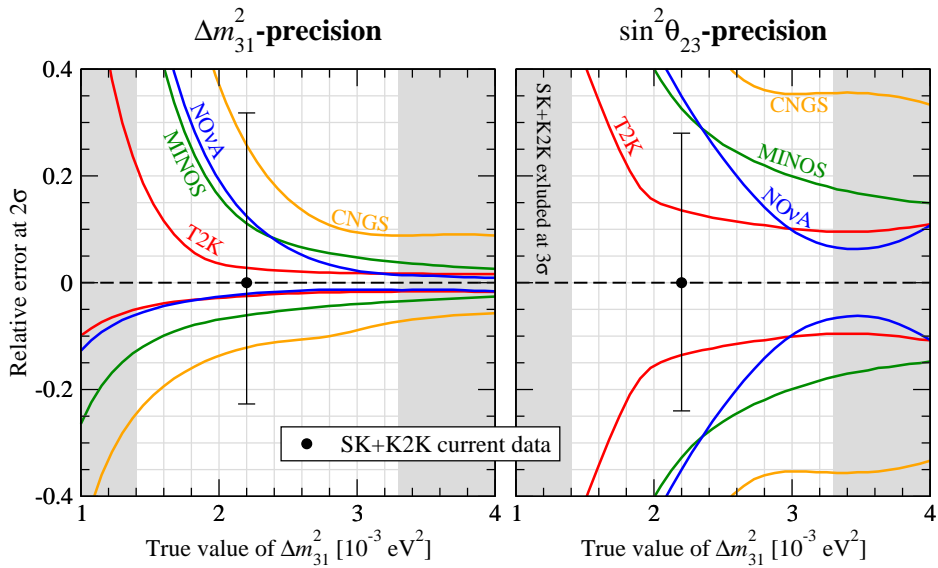}
\caption{The precision of $\Delta m^2_{31}$ (left panel) 
  and $\sin^2\theta_{23}$ (right panel) as a function of the
  ``true value'' of $\Delta m^2_{31}$ for $\theta_{23}^\mathrm{true} = \pi/4$
  (from Ref.~\protect~\cite{Huber:2004dv}).}
\label{fig:atm}
\end{figure}

Tab.~\ref{tab:atm} depends on the value of $\Delta m^2_{31}$ which
is shown in fig.~\ref{fig:atm}. The sensitivity suffers for all experiments 
for low values of $\Delta m^2_{31}$. T2K will provide a precise
determination of $\Delta m^2_{31}$ at the level of a few percent
for $\Delta m^2_{31} \gtrsim 2\times 10^{-3}$~eV$^2$. Although \nova\ 
can put a comparable lower bound on $\Delta m^2_{31}$, the upper bound is
significantly weaker, and similar to the bound from MINOS. The reason for this
is a strong correlation between $\Delta m^2_{31}$ and $\theta_{23}$, which
disappears only for $\Delta m^2_{31} \gtrsim 3\times 10^{-3}$~eV$^2$.
From the right panel of fig.~\ref{fig:atm} one can see that for $\Delta
m^2_{31} \sim 2\times 10^{-3}$~eV$^2$ only T2K is able to improve the current
bound on $\sin^2\theta_{23}$. One reason for the rather poor performance on
$\sin^2\theta_{23}$ is the fact that these experiments are sensitive mainly to
$\sin^22\theta_{23}$. This implies that for $\theta_{23} \approx \pi/4$ it is
very hard to achieve a good accuracy on $\sin^2\theta_{23}$, although
$\sin^22\theta_{23}$ can be measured with relatively high
precision~\cite{minakata}.

\begin{figure}[!ht]
\vspace*{3mm}
\includegraphics[width=7.9cm]{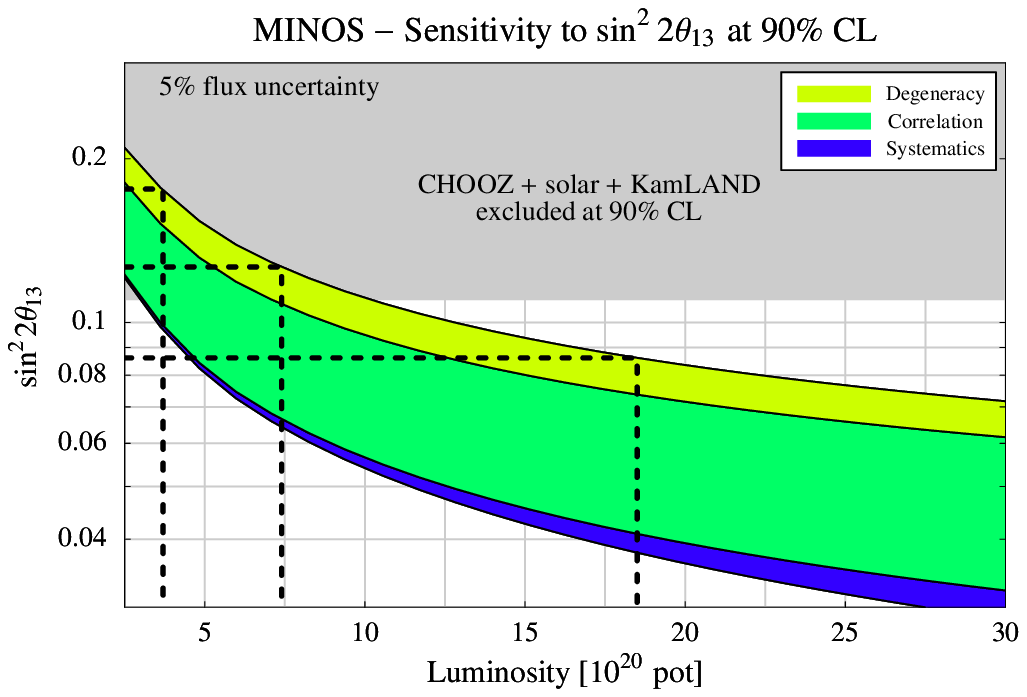}%
\hspace*{5mm}
\includegraphics[width=7.1cm]{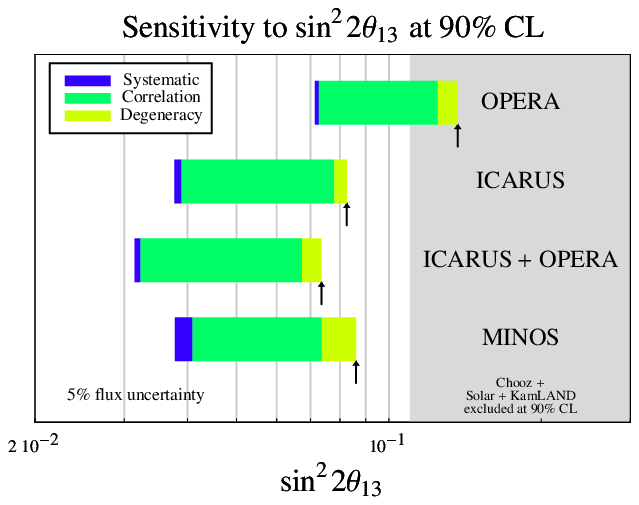}
\caption{Left plot: The sensitivity of the MINOS experiment 
to $\theta_{13}$ as a function of the protons on target (pot)
assuming a 5\% flux uncertainty. The dashed lines 
represent what 1,2 and 5 years of operation might achieve 
(from left to right). Right plot: Comparison of 5 years of 
operation for the MINOS and CNGS experiments. The grey area for
large $\sin^2 2\theta_{13}$ indicates in all cases the current 
limit from the CHOOZ experiment. The color code of the error
bars is explained in fig.~\protect\ref{fig:ferrors}.
Further details can be found in \protect~\cite{Huber:2004ug}.}
\label{fig:f1}
\end{figure}

Another interesting question is if the next generation 
long baseline experiments which will operate in the 
next years will be able to test the three flavourness 
of the oscillations. The sensitivity to a finite value 
of the key parameter $\theta_{13}$ is shown in 
fig.~\ref{fig:f1} for MINOS, OPERA and ICARUS. It can be
seen that these experiments have only a modest potential 
for improvements of the existing $\theta_{13}$ limit. 
%

\begin{figure}[tbh]
\centering
\includegraphics[width=0.5\textwidth]{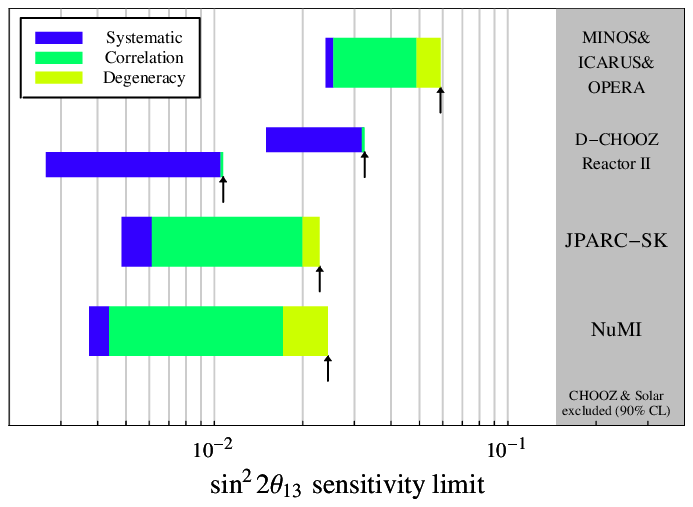}%
\hspace*{5mm}
\includegraphics[width=0.4\textwidth]{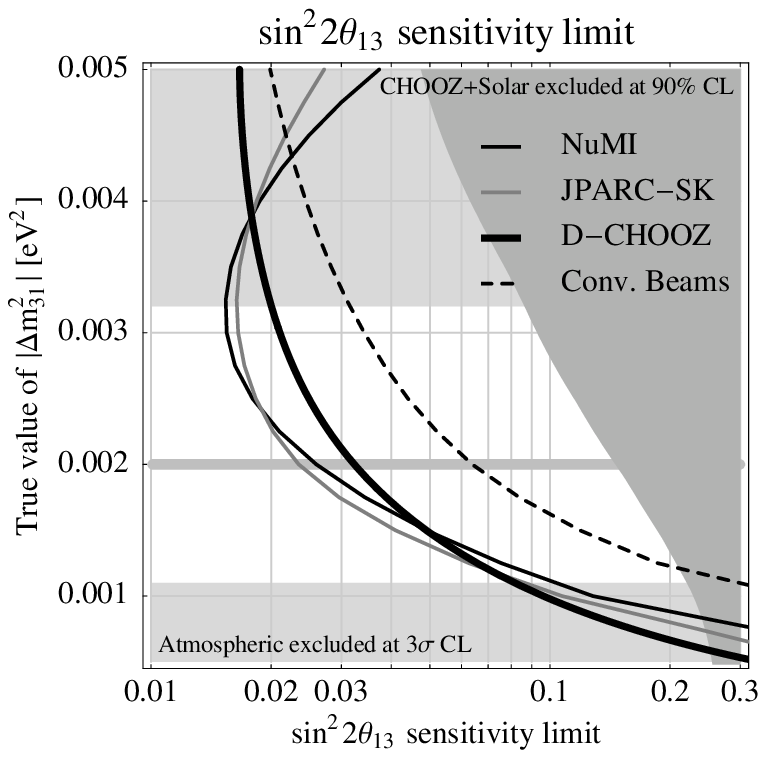}
\caption{Left plot: 
Sensitivity to $\stheta$ at 90\% CL for 
$\Delta m^2_{31} = 2\times 10^{-3}$~eV$^2$, 
$\Delta m^2_{21} = 7\times 10^{-5}$~eV$^2$.
Right plot: Sensitivity to $\stheta$ at 90\%~CL as a function of 
the ``true value'' of $\Delta m^2_{31}$.}
\label{fig:th13bars}
\end{figure}

The combined sensitivity to $\stheta$ of the next-to-next 
generation experiments is compared in the left panel of 
fig.~\ref{fig:th13bars} with new reactor experiments, 
T2K (JPARC-SK) and \nova\ (NuMI). 
It can be seen that the $\stheta$-limits from beam experiments 
are strongly affected by parameter correlations and degeneracies, 
whereas reactor experiments provide a ``clean'' measurement 
of $\stheta$, dominated by statistics and systematics~\cite{reactor}. 

The dependence of the $\stheta$-limit on the value of $\Delta m^2_{31}$
is shown in the right panel of fig.~\ref{fig:th13bars}, where the sensitivity of 
all experiments gets again rather poor for low values of $\Delta m^2_{31}$.  For
$\Delta m^2_{31} \sim 2\times 10^{-3}$~eV$^2$ we find roughly an improvement
by a factor 2 from conventional beam experiments (MINOS+ICARUS+OPERA
combined), a factor 4 from D-CHOOZ, and a factor 6 from the superbeams T2K and
\nova\ with respect to the current bound from global data~\cite{fit}. Note
that an optimized reactor experiment such as Reactor-II has the potential for
even better $\stheta$-sensitivities than the superbeams (c.f. left panel of 
fig.~\ref{fig:th13bars}).

\subsection{Synergies}

The previous discussion shows that competing plans with 
similar potential might be realized at the same time scale. 
This allows to combine the statistics of similar experiments
leading to improved global fits. However, it is also
possible to utilize synergies between experiments which 
are more than the simple addition of statistics. The point 
is that individual experiments measure a certain parameter 
combinations which contain different degeneracies and 
correlations. Experiments with similar sensitivities and
different correlations and degeneracies allow to separate 
the parameters partly or fully. An example of 
such a discussion is given by combining the T2K and \nova\ experiments 
for a fixed time of operation in the best possible way. T2K is 
essentially insensitive to matter effects, while matter effects 
play already some role for the longer \nova\ baselines. Both experiments 
could run partly with neutrino and partly with anti-neutrino beams. 
The cross-sections for anti-neutrinos are, however, smaller, leading to 
fewer events for the same running period. An anti-neutrino running is 
moreover in many aspects like a different experiment, but it is clear 
that anti-neutrino information is crucial in order to resolve the parameters.
A comparable reactor neutrino experiment would be very useful here.
It could provide the required information such that both T2K and
\nova\ could initially run fully in neutrino mode. 

Such a synergetic combination would be especially interesting if 
$\stheta$ would be close to the current bound. In order to demonstrate
these synergy effects we assume that $\stheta = 0.1$ and investigate 
what we could be learned about the CP-phase $\delta$ and the neutrino 
mass ordering. First we note that T2K, \nova\ and the reactor experiment 
will all be able to establish the non-zero value of $\stheta$.  However, depending 
on the unknown value of $\delta$ different values of $\stheta$ will be 
allowed. This can be seen as allowed regions in the $\theta_{13}$-$\delta$-plane 
shown in Figs.~8 and 9 of Ref.~\cite{Huber:2004ug}. None of the experiments 
on their own can give any information on the CP-phase $\delta$ and on 
the mass hierarchy. The determination of $\stheta$ from beam experiments 
is strongly affected by the correlations with $\delta$, and especially 
for \nova\ also correlations with other parameters are important. 
Moreover, the inability to rule out the wrong mass hierarchy leads 
to a further ambiguity in the determination of $\stheta$. In contrast, 
since the $\bar\nu_e$-survival probability does not depend on $\delta$, 
Reactor-II provides a clean determination of $\stheta$ at the level 
of 20\% at 90\%~CL. If all experiments are combined the complementarity 
of reactor and beam experiments allows to exclude up to 40\% of all 
possible values of the CP-phase for a given hierarchy. The wrong hierarchy 
can be ruled out at a modest confidence level with 
$\Delta\chi^2 \simeq 3$ due to matter effects in \nova. However, at high 
confidence levels all values of $\delta$ are allowed, and moreover, 
even for a given hierarchy CP-conserving and CP-violating values of 
$\delta$ cannot be distinguished at 90\%~CL. These results depend also 
to some extent on the value of $\delta$.

\begin{figure*}[tbh]
\centering
\includegraphics[width=0.65\textwidth]{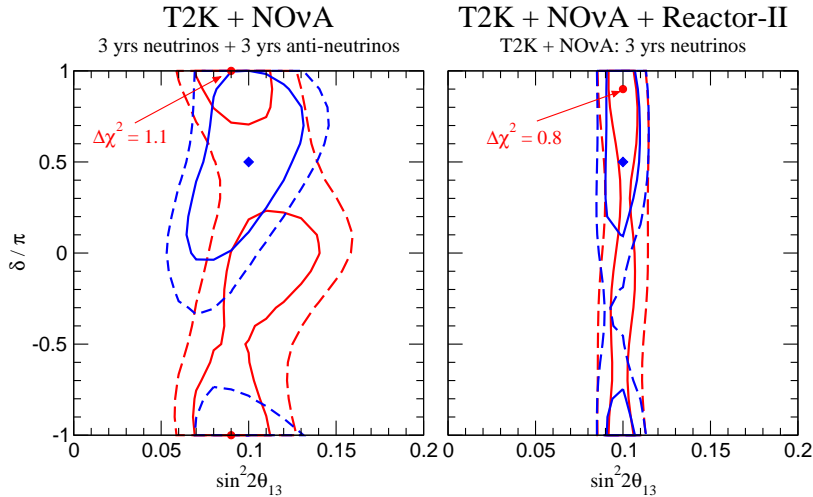}
\caption{Antineutrino running vs Reactor-II. We show the 90\%~CL (solid
   curves) and $3\sigma$ (dashed curves) allowed regions in the
   $\stheta$-$\delta$-plane for the assumed values $\stheta=0.1$ and
   $\delta=\pi/2$. The blue/dark curves refer to the allowed regions for
   the normal mass hierarchy, whereas the red/light curves refer to the
   $\mathrm{sgn}(\Delta m^2_{31})$-degenerate solution (inverted hierarchy),
   where the projections of the minima onto the $\stheta$-$\delta$-plane are
   shown as diamonds (normal hierarchy) and dots (inverted hierarchy). For the
   latter, the $\Delta\chi^2$-value with respect to the best-fit point is also
   given.}
\label{fig:anti-vs-react}
\end{figure*}

So far we have considered only neutrino running for the superbeams, since it
is unlikely that significant data can be collected with anti-neutrinos within
ten years from now. Nevertheless, it might be interesting to investigate the
potential of a neutrino-antineutrino comparison. In
fig.~\ref{fig:anti-vs-react} we show the results from T2K+\nova\ with 3 years 
of neutrinos plus 3 years of anti-neutrinos each (left), in comparison with the case
where the antineutrino running is replaced by Reactor-II (right). We find that
antineutrino data at that level does neither solve the problems related to the
CP-phase nor to the hierarchy. Still CP-violating and CP-conserving values
cannot be distinguished at 90\% CL. Moreover, the determination of $\stheta$
is less precise than from the reactor measurement. To benefit from
antineutrino measurements a significantly longer measurement period would be
necessary, to obtain large enough data samples.

\subsection{Long term perspectives}

\begin{figure}[!ht]
\includegraphics[width=7.5cm]{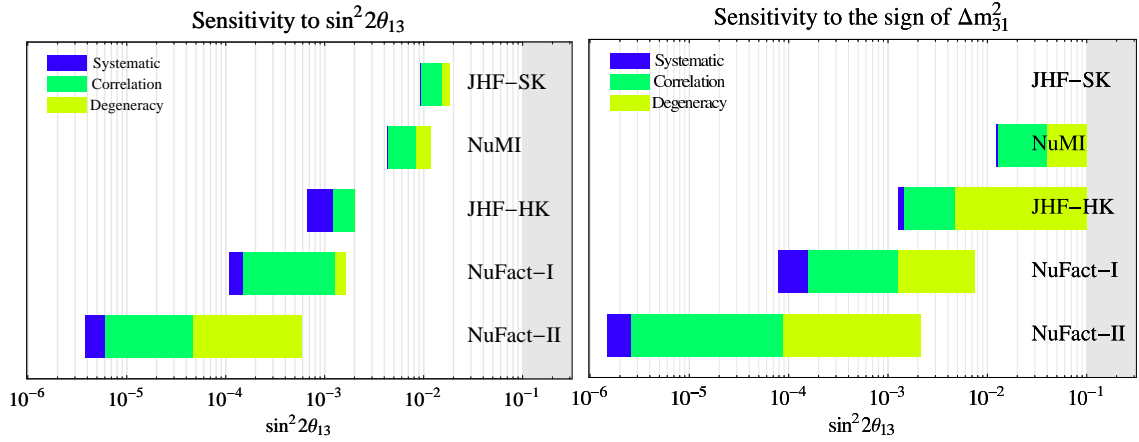}%
\includegraphics[width=7.5cm]{./signbar2.eps}
\caption{Left plot: The $\theta_{13}$ sensitivity of different future 
accelerator based neutrino oscillation experiments 
\protect~\cite{Huber:2002mx}.
Right plot: The $\theta_{13}$ values for which sensitivity to
matter effects, i.e. $sign(\dm{31})$ exists. The shown bands 
are again the reduction of sensitivity from a purely statistical 
limit (left end of the dark grey/blue range) by systematics 
(right end of dark grey/blue), correlations (medium grey/green) 
and degeneracies (light grey/yellow). The right end of the light 
grey(yellow) band represents the final 90\%CL limit. The grey 
area for large $\sin^2 2\theta_{13}$ indicates the current 
limit from the CHOOZ experiment.}
\label{fig:f2}
\end{figure}

Beyond the discussed accelerator and reactor based oscillation 
experiments exist more ambitious projects like the JHF-HyperKamiokande 
project, beta beams and neutrino factories. Such experiments clearly 
require further R\&D before they can be built. However, assuming 
current knowledge, we believe that such setups are possible in the 
long run. The potential of the JHF-HyperKamiokande experiment 
and a neutrino factory are compared in fig.~\ref{fig:f2} to T2K 
(JHF-SK) and \nova\ (NuMI). It can be seen that the existing limits 
can be improved by a few orders of magnitude compared to now. 


\section{Theoretical implications and conclusions}

One of the most interesting unsolved topics is the origin of
flavour and fermion masses. There exist apparent regularities 
in the fermionic field content which make it very tempting to 
introduce right-handed neutrino fields leading to both Dirac 
and Majorana mass terms for neutrinos. Diagonalization of the 
resulting mass matrices yields Majorana mass eigenstates and 
due to the see-saw mechanism very small neutrino masses. This 
can also be nicely realized in embeddings of the SM into GUTs 
with larger symmetries, such as SO(10). Before the discovery 
of large leptonic mixing, many theorists expected the leptonic
mixings to be similar to quark mixing, characterized by small 
mixing angles. Experiment led theory in showing the striking 
results that $\sin^2 2\theta_{23} \simeq 1$ and 
$\tan^2 \theta_{12} \simeq 0.39$, while $\theta_{13}$ is small. 
By finding two large mixing angles, neutrino physics has already 
provided surprising and very valuable information which severely 
constrains models of neutrino masses. Future precision neutrino 
oscillation experiments will provide further precision tests of 
the flavour sector. The level of precision will confirm or rule 
out ideas about the origin of flavour and connected topics.

An important subject is the small value of $\sin^2 2\theta_{13}$. Since there
are two large mixing angles, there is no particular reason to 
expect the third angle, $\theta_{13}$, to be extremely small or even 
zero. A small value of $\stheta\simeq 0.1$ could be a numerical 
coincidence in a framework which predicts generically large or 
sizable mixings. However, if the limit on $\stheta$ would become 
smaller by an order of magnitude then some protective 
mechanism like a ``symmetry'' would be required. 
This can be seen in neutrino mass models which are able to 
predict the large values for $\theta_{12}$ and $\theta_{23}$. 
Such models have a certain tendency to predict a sizable value 
of $\theta_{13}$ as can be seen, for example, 
in~\cite{Barr:2000ka,Altarelli:2002hx,Barbieri:2003qd,Chen:2003zv,King:2003jb,Goh:2003hf}.
The conclusion is that a value of $\theta_{13}$ close to the CHOOZ 
bound would be quite natural, while much smaller values are less
likely or hard to understand.

Future precision measurements can also test if relations like 
$\theta_{23}=\pi/4$ \cite{Antusch:2004yx} or $\theta_{12}+\theta_{C}=\pi/4$ 
which are currently fulfilled within experimental errors still hold at 
much better precision. If so, then this would provide strong 
constrains on the origin of the flavour structure. 
Precision is also valuable, even if such special relations are 
not found. The point is that it is generally not easy to predict
a set of very precisely known masses and mixings in a certain 
model or class of models. 

Neutrino masses and mixing parameters are also subject to quantum 
corrections between low scales, where measurements are performed, 
and high scales where some theory typically predicts the masses 
and mixings. This has interesting implications, since it implies
that certain deviations from special relations are expected due 
to quantum corrections (renormalization group or RGE effects). 
Suppose, for example, that some theory were able to predict 
$\theta_{13}\equiv 0$. Then RGE effects would still predict a 
tiny, but finite value at low energy. Strictly speaking, 
$\theta_{13}=0$ cannot be excluded completely by this argument, 
as the high-energy value could be just as large as the change
due to running and of opposite sign. However, a complete 
cancellation of this kind would be a miraculous fine-tuning, 
since the physics generating the value at high energy is not 
directly related to quantum corrections at lower energies. 
The strength of the running of $\theta_{13}$ depends on the 
neutrino mass spectrum and whether or not supersymmetry is realized.
For the Minimal Supersymmetric Standard Model one finds a shift 
$\Delta \sin^2 2\theta_{13} > 0.01$ for a considerable parameter range,
i.e.\ one would expect to measure a finite value of $\theta_{13}$
~\cite{Antusch:2003kp}.  Conversely, limits on model parameters would 
be obtained if an experiment were to set an upper bound on 
$\sin^2 2\theta_{13}$ in the range of 0.01. In any case, it should be 
clear that a precision of the order of quantum corrections to neutrino 
masses and mixings is very interesting in a number of ways.

Precision measurements would also allow very interesting tests of many 
other topics, like MSW matter effects, three neutrino unitarity, 
neutrino decay, de-coherence and NSI effects. There exists also 
an interesting interplay with theories beyond the Standard Models, 
flavour models as well as astroparticle physics (leptogenesis, supernovae, 
nucleosynthesis, structure formation). In summary, future precision 
oscillation experiments will provide the best window into the so 
far un-understood flavour sector. It may give us a glimpse on the 
origin of flavour, but it may also lead to unexpected results and 
insights as it happened before in neutrino physics.

\bigskip\noindent
{\large \bf Acknowledgments:}
Work supported by the ``Sonderforschungsbereich 375 
f{\"u}r Astro-Teilchenphysik der Deutschen For\-schungs\-ge\-mein\-schaft''.


\end{document}